# The spectrum of Sgr A* and its variability

*Dedicated to Prof. Peter G. Mezger on the occasion of his 65th birthday*

**Wolfgang J. Duschl**[1,2,⋆] **and Harald Lesch**[1]

[1] Max-Planck-Institut für Radioastronomie, Auf dem Hügel 69, Postfach 2024, D-53010 Bonn, Germany
[2] Institut für Theoretische Astrophysik, Universität Heidelberg, Im Neuenheimer Feld 561, D-69120 Heidelberg, Germany

**Abstract.** We demonstrate that there is only one physical process required to explain the spectrum and the variability of the radio source at the dynamical center of our Galaxy, Sgr A*, in the frequency range from ≈1 to ≈1000 GHz, namely optically thin synchrotron radiation that is emitted from a population of relativistic electrons. We attribute the observed variability to variable energy input from an accretion disk around Sgr A* into the acceleration of the electrons.

## 1. Introduction

Some years ago Lesch et al. (1988) showed that the nonthermal radio spectra of the Galactic Centre, including Sgr A* and the extended components (often referred to as *Bridge* and *Arc*, Brown & Liszt 1984, Reich et al. 1988), can be explained as optically thin synchrotron radiation due to a monoenergetic electron distribution. Then it was *assumed* that Sgr A* is the source of the energetic particles, which propagate into the extended components. Now, due to increasing resolution, the spectrum of Sgr A* itself can be traced up to $10^{13}$ Hz. For frequencies up to at least $10^{12}$ Hz Sgr A* exhibits the same inverted spectrum as the Bridge and the Arc, namely $S_\nu \propto \nu^{1/3}$ ($\nu$: frequency, $S_\nu$: flux density) (see sect. 2 and references there). This is the typical spectrum of electrons following a monoenergetic relativistic distribution and radiating optically thin synchrotron radiation. Thus, we can now make a much stronger point in favour of Sgr A* being a source of monoenergetic electrons, indeed.

In the following first we discuss measurements of the spectrum of Sgr A* and its time variability (Sect. 2). Then (Sect. 3) we give arguments which favour the existence of a monoenergetic electron distribution in the vicinity of an accreting black hole and calculate (in Sect. 4) the energetics of such a radiating system. In Sect. 5 we discuss the amplification of the required magnetic fields in an accretion disk around a central black hole. Finally (Sect. 6) we demonstrate its applicability to the Galactic Centre and to Sgr A* in particular.

## 2. The spectrum of Sgr A* and its variability

The spectrum of Sgr A* is described by Zylka et al. (1992) and by Mezger (1994). Its variabilty is well established. In figure 1 we summarize the range of observed variabilty from 1.5 GHz to 86 GHz (full lines), and individual measurements at higher frequencies up to 375 GHz (when given in the literature, broken lines show the error bars of the individual measurement; otherwise an asterisk indicates the measured value). The observations are taken from Backer (1982, 1994), Zylka & Mezger (1988), Zhao et al. (1989, 1991), Zhao & Goss (1993), Zylka et al. (1992), and Dent et al. (1993).

Additionally, in figure 1 we give two best fit lines (linear in log $S_\nu$ vs. log $\nu$). The steeper one represents all observations given in the figure. Its dependency is $S_\nu \sim \nu^{0.38}$. For the shallower one we took into account only the frequencies for which a variation is well established (full lines). This ensures that we exclude the chance that a single observation at a certain frequency accidentally gives an extremal value within its range of possible variability. For the latter representation we get a dependency of $S_\nu \sim \nu^{0.35}$.

## 3. A monoenergetic electron distribution

The particle momentum or energy distribution function is the result of the interplay of energy gain and energy loss processes. Acceleration due to diffusive shock waves, magnetohydrodynamical (MHD) waves or magnetic reconnection shift the particles upwards in energy space, whereas radiation processes like synchrotron or inverse Compton scattering shift the particle towards lower energies. In a steady-state situation the resulting

---

*Send offprint requests to*: W.J. Duschl, Heidelberg
⋆ E-Mail: wjd@platon.ita.uni-heidelberg.de

**Fig. 1.** The range of observed variabilty of Sgr A* from 1.5 GHz to 86 GHz (full lines), individual measurements at higher frequencies up to 375 GHz (when given in the literature, broken lines show the error bars of the individual measurement; otherwise an asterisk indicates the measured value). The observations are taken from Backer (1982, 1994), Zylka & Mezger (1988), Zhao et al. (1989, 1991), Zhao & Goss (1993), Zylka et al. (1992), and Dent et al. (1993).

equilibrium distribution function of relativistic electrons in the momentum range $p > p_M/a$ is given by (Lesch and Schlickeiser 1987)

$$f(p) \simeq p^a \exp(-p/p_M). \qquad (1)$$

$a$ denotes the ratio of the rate of shock wave acceleration (or due to magnetic reconnection) to the rate of acceleration by MHD waves. $p_M$ denotes that momentum where the timescale for diffusive shock wave acceleration equals the timescale for inverse Compton and synchrotron losses. Equation (1) holds provided the escape time of particles $t$ is much longer than the acceleration time, i.e., for a (quasi-)stationary situation. In the case of magnetic reconnection the maximum particle energy is given by the product of the length of the reconnection sheet and the electric field parallel to the magnetic field lines (Lesch and Reich 1992; Lesch and Pohl 1992). Strictly speaking, a monoenergetic distribution at $p = p_M$ will be maintained if

$$\dot{p}_{\text{Gain}}(p_M) + \dot{p}_{\text{Loss}}(p_M) = 0. \qquad (2)$$

The physical mechanism for the accumulation of particles around $p_M$ in the case of shock wave and MHD wave acceleration is evident: at momenta $p < p_M$ momentum gain dominates momentum loss, $\dot{p}_{\text{Gain}} + \dot{p}_{\text{Loss}} > 0$ and the electrons are convected upwards in momentum space. At momenta $p > p_M$ the opposite is true $\dot{p}_{\text{Gain}} + \dot{p}_{\text{Loss}} < 0$ and the electrons are convected downwards in momentum space. Without momentum diffusion the equilibrium distribution would be singular at $p = p_M$, because there would be no physical process to remove particles from $p = p_M$. Since we allowed momentum diffusion of particles this $\delta$-function behaviour is smoothed out and an isotropic monoenergetic distribution function is formed.

This argumentation is also applicable to acceleration by magnetic reconnection. The approaching magnetic field lines confine the particles in a current sheet. These particles can only escape along the field lines, where the electric field accelerates much slower than the motion along the field lines. Thus, the escape time is longer than their acceleration time and results in a monoenergetic electron distribution function, i.e., as a relativistic electron beam (Lesch and Reich 1992). Relativistic electron beams are unstable with respect to the excitation of plasma waves. The waves isotropize the beam electrons and the initially anisotropic distribution function develops into an isotropic monoenergetic one. This isotropisation happens on the timescale (Achatz et al. 1990)

$$t_{\text{iso}} \simeq \omega_{\text{pe}}^{-1} \frac{n_{\text{e}}}{n_{\text{r}}} \left[ \frac{m_{\text{p}}}{m_{\text{e}}} \right]^{1/2}. \quad (3)$$

$\omega_{\text{pe}} \simeq 5.6 \cdot 10^4 \sqrt{n_{\text{e}}}$ is the electron plasma frequency, $n_{\text{e}}$ is the number density of thermal electrons, $n_{\text{r}}$ is the number density of relativistic electrons and $m_{\text{p}}$ ($m_{\text{e}}$) is the proton (electron) mass.

In the following we assume that the isotropisation time is shorter than the energy loss time due to radiation and we will check this assumption a posteriore.

In both acceleration scenarios (shocks and MHD waves or magnetic reconnection) an isotropic monoenergetic electron distribution appears. Such a distribution produces an inverted optically thin synchrotron spectrum with a spectral index around $+1/3$ which is in good agreement with the observed time averaged spectrum of Sgr A* for frequencies up to $10^{12}$ Hz.

## 4. The energetics of a radiating plasma with monoenergetic relativistic electrons

In the following, we discuss the energetics of a radiating plasma with monoenergetic relativistic electrons, and use the observed luminosity $L$ of Sgr A* of a few times $10^4 L_\odot$ (Mezger 1994, and references therein). The luminosity $L$ of an object with emission volume $V$ filled with radiating monoenergetic relativistic electrons is given by (Lang 1978)

$$L \simeq P(\gamma) n_{\text{r}} V, \quad (4)$$

where

$$P(\gamma) = 1.6 \cdot 10^{-15} B^2 \gamma^2 \text{ erg s}^{-1} \quad (4a)$$

denotes the emitted power of one electron with the Lorentz factor $\gamma$ emitting synchrotron radiation in a magnetic field $B$.

We estimate the number density of relativistic electrons by assuming equipartition between energy density of relativistic electrons and magnetic energy density:

$$n_{\text{r}} \simeq \frac{B^2}{8\pi \gamma m_{\text{e}} c^2}. \quad (5)$$

The emission volume is taken from the 86 GHz observations of Krichbaum et al. (1993) who observed a source size of about 0.34 milliarcseconds, which correspond to a linear scale of $4.42 \cdot 10^{13}$ cm at a distance of the galactic centre of 8.5 kpc. If the emission region is an elongated structure, as indicated by the VLBI measurements (for details of the structure, see Krichbaum et al. 1993) the volume is $V \simeq 2.7 \cdot 10^{41}$ cm$^3$.

form the formula for synchrotron radiation. For a monoenergetic distribution function the typical emitted frequency corresponds to the maximum frequency $\nu_{\text{max}}$ at which the spectrum exponentially cuts off (see fig. 2).

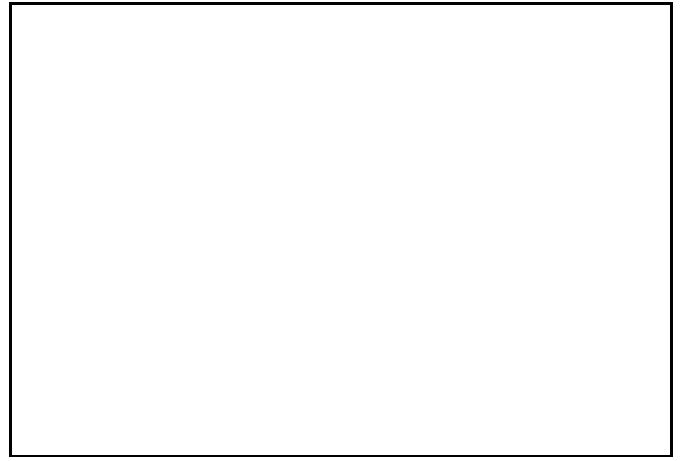

**Fig. 2.** The optically thin synchrotron spectrum of an ensemble of relativistic monoenergetic electrons (full line). The broken line demonstrates the quality of the approximation with $S_\nu \sim \nu^{1/3}$ for frequencies below the cut-off.

$$\nu_{\text{max}} \simeq \gamma^2 \nu_{\text{ce}}, \quad (6)$$

where the electron gyrofrequency is given by $\nu_{\text{ce}} = \frac{eB}{2\pi m_e c}$. Thus, the particle Lorentz factor is

$$\gamma \simeq \sqrt{\frac{2\pi m_e c \nu_{\text{max}}}{eB}}. \quad (7)$$

Inserting (4a), (5), and (7) into 4 we obtain

$$L \simeq 4.65 \cdot 10^{-14} \nu_{\text{max}}^{1/2} B^{7/2} V, \quad (8)$$

which results in a magnetic field strength

$$B \simeq 13.8 \, \text{G} \left[ \frac{\nu_{\text{max}}}{10^{12} \, \text{Hz}} \right]^{-1/7} \times$$

$$\times \left[ \frac{R}{4.42 \cdot 10^{13} \, \text{cm}} \right]^{-6/7} \left[ \frac{L}{10^4 \, L_\odot} \right]^{2/7}. \quad (9)$$

With Eq. (9) we obtain from Eq. (7) the typical Lorentz factor of the radiating electrons

$$\gamma \simeq 162 \left[ \frac{\nu_{\text{max}}}{10^{12} \, \text{Hz}} \right]^{1/2} \left[ \frac{B}{13.8 \, \text{G}} \right]^{-1/2}, \quad (10)$$

which corresponds to an energy of about 250 MeV.

The energy loss time of these particles is given by

$$t_{\text{Loss}} \simeq \frac{5 \cdot 10^8}{\gamma B^2} \simeq 0.2 \, \text{days} \left[ \frac{\gamma}{162} \right]^{-1} \left[ \frac{B}{19.3 \, \text{G}} \right]^{-2}. \quad (11)$$

electrons from Eq. (5)

$$n_\text{r} \simeq 5.71 \cdot 10^4 \text{ cm}^{-3} \left[\frac{\gamma}{162}\right]^{-1} \left[\frac{B}{19.3\,\text{G}}\right]^2. \quad (12)$$

Next, we check whether the isotropisation time [Eq. (3)] is shorter than $t_\text{Loss}$ as we assumed in the beginning. According to Eq. (3) $t_\text{iso} \simeq 7.5 \cdot 10^{-4}$ sec $n_\text{e}^{1/2}/n_\text{r}$. Since Sgr A$^*$ is embedded in a HII-region, named Sgr A West (Lacy et al. 1980; 1982) we choose as an example the density of a compact HII-region $n_\text{e} \simeq 10^6$ cm$^{-3}$. We obtain a fraction of a second for $t_\text{iso}$, which is much shorter than any reasonable energy loss time [eq. (11)]. Hence our assumption of an isotropic monoenergetic electron distribution is exceedingly likely.

## 5. Accretion disk and magnetic field amplification

Finally, we have to check whether our result for the magnetic field strength is reasonable in terms of an accretion disk model, including a massive black hole of the order of $10^6 M_\odot$ in its center.

Magnetic fields have long been considered an important element in the dynamics of accretion disks, primarily as a mechanism for supplying internal stresses required for efficient angular momentum transfer (Eardley and Lightman 1975; Ichimaru 1977). According to these models magnetic fields are amplified within the inner portion of an accretion disk by the joint action of thermal convection and differential rotation along Keplerian orbits. Field amplification will then be limited by nonlinear effects; as a consequence of buoyancy, magnetic flux will be expelled from the disk, leading to an accretion disk corona consisting of many magnetic loops where the energy is stored and probably transferred via magnetic reconnection into heat and particle acceleration, leading to a state in which loops contain very hot, relatively low-density plasma (compared to the disk).

Magnetic field amplification due to differential motion of a conductive disk medium is usually described in the MHD limit through the induction equation for the magnetic field (cf. Parker 1979)

$$\frac{\partial \mathbf{B}}{\partial t} = \nabla \times [\mathbf{v} \times \mathbf{B}] + \frac{c^2}{4\pi\sigma}\nabla^2 \mathbf{B}, \quad (13)$$

where $\mathbf{B}$ is the magnetic field induction, $\mathbf{v}$ the velocity of the fluid, and $\sigma$ the electrical conductivity. Due to the high temperatures in the disk (up to $4 \cdot 10^4\,K$; Zylka et al. 1992) the electrical conductivity is high. Therefore, magnetic field evolution via the plasma differential motion is well described by Eq. (13) even if we neglect the classical (collisional) dissipation of the magnetic field implied by the last term on the right-hand side of Eq. (13). In the case of Keplerian differential rotation, Eq. (13) describes (assuming an axisymmetric geometry) the amplification of a toroidal magnetic field $B_\text{T}$ in the presence of a seed poloidal magnetic field $B_\text{P}$:

$$\frac{\partial B_\text{T}}{\partial t} = R\frac{\partial \Omega}{\partial R}B_\text{P}, \quad (14)$$

material about the black hole with mass $M_\text{BH}$.

The buildup of magnetic fields within the disk is limited by nonlinear effects related to convection. Since convection takes place primarily perpendicular to the plane of the disk, we shall assume differential rotation to remain the dominant mechanism for the toroidal magnetic field $B_\text{T}$ amplification; the amplification of the poloidal field $B_\text{P}$ will then be dominated by convection-mediated effects. For convection cells whose aspect ratio is $\sim 1$, the poloidal magnetic field spatial scale will then be of the order of the convective cell size $H$, equal to the half-thickness of the disk. To describe the amplification of $B_\text{P}$ we invoke magnetic flux conservation:

$$B_\text{P} \simeq \frac{H}{R}B_\text{T}. \quad (15)$$

The maximal toroidal field strength is given by the energy density of the turbulence in the disk (Galeev et al. 1979)

$$\frac{B_\text{T}^2}{8\pi} \simeq n_\text{e} m_\text{p} c_\text{s}^2, \quad (16)$$

where $c_\text{s} = \sqrt{\frac{k_\text{B}T}{m_\text{p}}}$ denotes the sound velocity.

The effective temperature in the accretion disk at a radius $R$ from the centre of the accreting black hole $T_\text{accr}(R)$ is given by (see, e.g., Frank, King & Raine 1992)

$$T_\text{accr}(R) \simeq T_\text{in}\left[\frac{R_\text{in}}{R}\right]^{0.75}, \quad (17)$$

with the temperature at the disk's inner edge $R_\text{in} \simeq 3R_\text{S}$ of

$$T_\text{in} = 7.1 \cdot 10^4\,\text{K} \left[\frac{M_\text{BH}}{10^6\,M_\odot}\right]^{1/4} \left[\frac{\dot{M}}{10^{-6}\,M_\odot\,\text{yr}^{-1}}\right]^{1/4}. \quad (18)$$

A temperature of 71,000 K corresponds to a sound velocity of $2.4 \cdot 10^6$ cm s$^{-1}$.

For eqs. (17) and (18), we use the following approximations: the disk is stationary (i.e., the radial mass flow rate $\dot{M}$ is constant); the non-relativistic limit of disk physics applies; and, the disk's inner edge can be described with the same physics as regions within the disk (i.e., we neglect the term coming from the radially inner boundary condition). It is known that the latter two approximations introduce only errors that are limited to the radially innermost disk regions and that correspond to factors of the order of 1, i.e., these approximations will not influence our conclusions.

With the same parameters we obtain for the density a central value of about $n_\text{e} \simeq 4 \cdot 10^{12}$ cm$^{-3}(R/R_\text{in})^{-3/2}$. Altogether we get from Eq. (18) and (16) the maximum toroidal field strength of about $B_\text{T} \simeq 30$ G or a poloidal field of about $B_\text{P} \simeq 1$ G, respectively.

Since the only available radio observation which can be used for innermost parts of Sgr A* is the VLBI measurement by Krichbaum et al. (1993) we have to deal with an elongated structure, which might be interpreted as a jet (Falcke et al. 1993). Hence, we will take the poloidal field as the field strength which make the relativistic electrons radiate synchrotron emission. A value of a few Gauss is very close to the value we obtained from Eq. (9).

Averaged over time, the spectrum of Sgr A* is very close to a $S_\nu \sim \nu^{1/3}$ law (Fig. 1 in sect. 2). But it is important to note that individual spectra vary on timescale of weeks (Zhao et al. 1989, 1991; Zhao & Goss 1993) and sometimes differ considerably from a $S_\nu \propto \nu^{1/3}$ dependency.

Together with the above shown consistency of the observed and the theoretical magnetic field, this leads us to the conclusion that what we observe in Sgr A* is indeed driven by a monoenergetic distribution of electrons emitting synchrotron radiation. The variability is then due to a variable energy input from the accretion disk. As shown above [eq. (11)] the energy loss time is of the order of a few days. Within such a time scale, the electron population can react upon a variable supply of energy for the acceleration process.

Thus, the fact that the time averaged spectrum indeed shows a $\nu^{1/3}$ dependency tells us that the timescale within which the energy input into the acceleration process may vary must be of the same order as the loss time scale.

Were the timescale of the input very much shorter then the electron distribution would "see" only the averaged energy input rate, i.e., we would not expect any variations of the spectrum at all. This argument again relys on an accretion rate that is constant on the average. It also takes not into account other reasons for variability (e.g., refractive interstellar scintillation; this process could account for [some] variability at the lower frequencies [$\leq 10$ GHz], but hardly for the higher ones; see, e.g., Zhao et al. 1990).

Were the time scale very much longer then again we would not expect deviations from the proportionalty to $\nu^{1/3}$ as then the electrons could follow the changes in a quasi-stationary fashion.

With the above estimate for the sound velocity, we find that characteristic viscosity time scales in the central regions of an $\alpha$ accretion disk (i.e., at radial distances of a few to several $R_S$) are of the order of a few days, i.e., of the same order as the energy loss time scale of the electrons. This demonstrates that the observed time scale of spectral variations and both the theoretically deduced time scales (electron energy loss and viscous changes in a disk) are in good agreement with each other.

## 7. Conclusions

We have shown that the spectrum of Sgr A* in the frequency range of $\log(\nu/\text{Hz}) \approx 9\ldots 12$ and its variability can be interpreted as being due to a single underlying physical mechanism, that emit synchrotron radiation.

For our arguments the variability of the spectrum is crucial. Only the near coincidence of the two relevant time scales (accretion time scale and energy loss time scale of the electrons) allows a variability in which individual spectra may very well deviate from the $\nu^{1/3}$ law while – on the average – the source *knows* about the steady state solution.

In a next step one has to look into the details of the time dependent evolution of such an electron distribution. This then will also allow to decide whether only the high frequency variations ($> 10$ Ghz) are due to intrinsic source variations. For lower frequencies, in our model, variability could also be due to source variability. On the other hand, Zhao & Goss (1993) give arguments that favour scattering of the radiation by the ISM along the line of sight as the reason for variability at the lower frequencies. It seems as if this can be clarified only by a time dependent model. But such a model is beyond the scope of our present work that is presented in this paper.

Supporting evidence for Sgr A* being a source of monoenergetic relativistic electrons comes from the fact that also the Bridge and the Arc in the Galactic Center show the same kind of spectrum (Lesch et al. 1988).

*Acknowledgements.* We benefitted much from discussing the subject with Prof. P.G. Mezger and Dr. R. Zylka. WJD acknowledges support from the *Bundesministerium für Forschung und Technologie* through a *Verbundforschungs*-grant (PH/055HD45A) and from the *Deutsche Forschungsgemeinschaft* through *Sonderforschungsbereich 328 "Evolution of Galaxies"*.

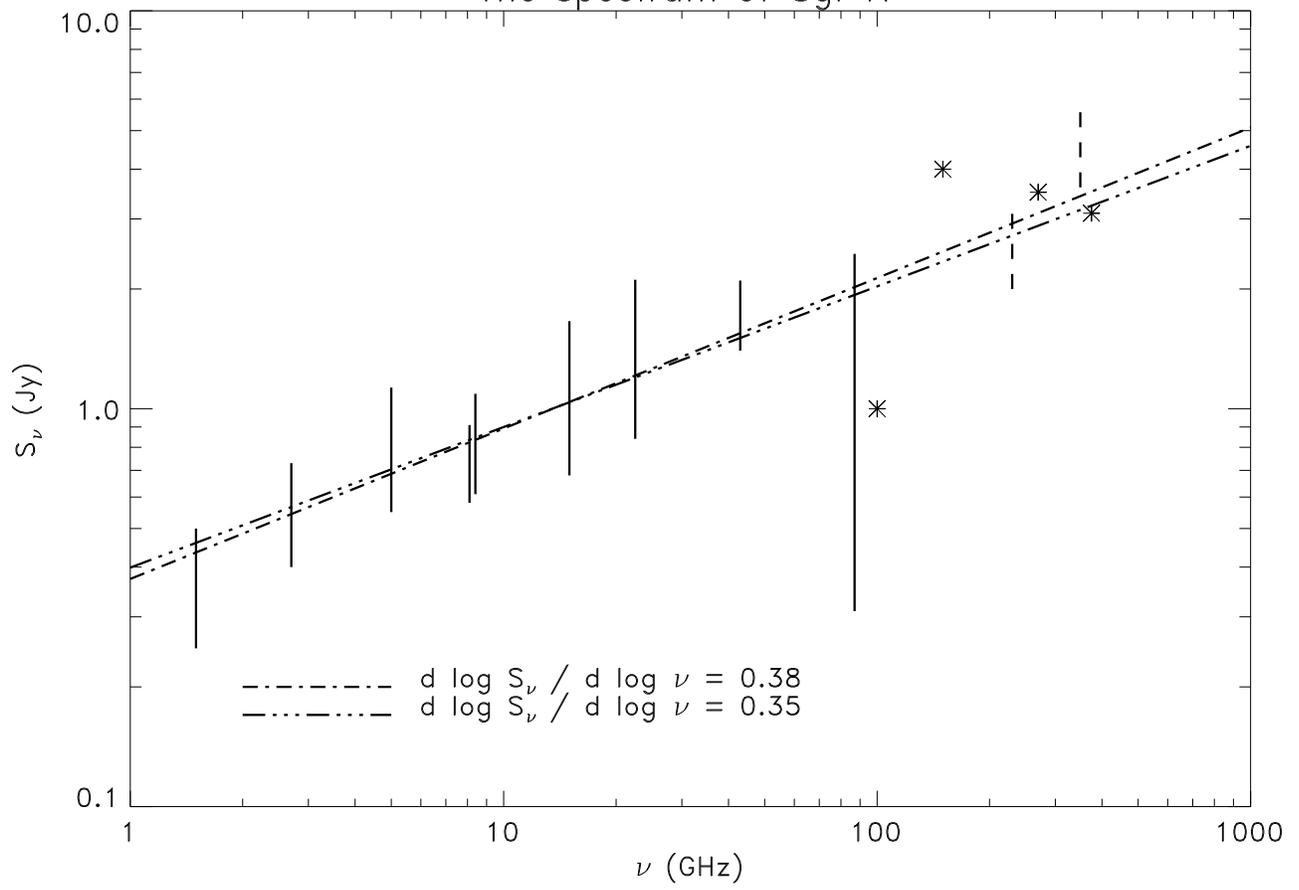

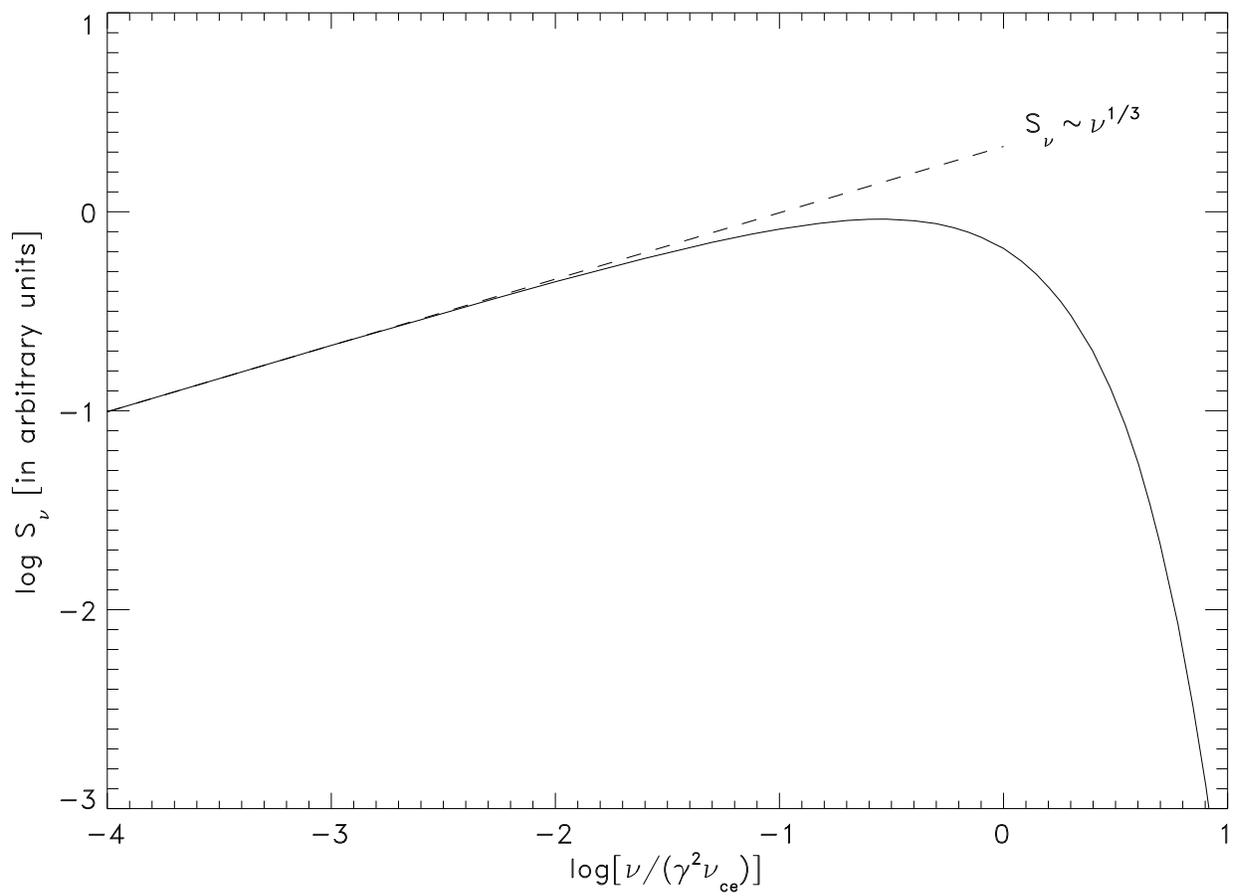